\documentclass[runningheads]{llncs}
\usepackage{graphicx}
\usepackage{comment}
\usepackage{hyperref}
\usepackage{wrapfig}
 \usepackage{url}
\usepackage[font=scriptsize,skip=2pt]{caption}

\begin{document}

\title{
SIMPT: Process Improvement Using Interactive Simulation of Time-aware Process Trees \thanks{\scriptsize{ Funded by the Deutsche Forschungsgemeinschaft (DFG, German Research Foundation) under Germany's Excellence Strategy – EXC 2023 Internet of Production- Project ID: 390621612. We also thank the Alexander von Humboldt (AvH) Stiftung for supporting our research.}}}
\titlerunning {SIMPT}
\author{Mahsa Pourbafrani\inst{1}\and 
Shuai Jiao\inst{2} \and
Wil M. P. van der Aalst\inst{1}}
\authorrunning{M. Pourbafrani et al.}
\institute{Chair of Process and Data Science, RWTH Aachen University, Germany  \email{\{mahsa.bafrani,wvdaalst\}@pads.rwth-aachen.de} \and RWTH Aachen University, Germany \\
 \email{ shuai.jiao@rwth-aachen.de}
 }

\maketitle              
\begin{abstract}
Process mining techniques including process discovery, conformance checking, and process enhancement provide extensive knowledge about processes. 
Discovering running processes and deviations as well as detecting performance problems and bottlenecks are well-supported by process mining tools. However, all the provided techniques represent the past/current state of the process. The improvement in a process requires insights into the future states of the process w.r.t. the possible actions/changes.  
In this paper, we present a new tool that enables process owners to extract all the process aspects from their historical event data automatically, change these aspects, and re-run the process automatically using an interface. The combination of process mining and simulation techniques provides new evidence-driven ways to explore "what-if" questions. 
Therefore, assessing the effects of changes in process improvement is also possible. 
Our Python-based web-application provides a complete interactive platform to improve the flow of activities, i.e., process tree, along with possible changes in all the derived activity, resource, and process parameters. 
These parameters are derived directly from an event log without user-background knowledge.


\end{abstract}
\section{Introduction}


The real value of providing insights by process mining emerges when these insights can be put into action \cite{DBLP:books/sp/Aalst16}. Actions include the improvement of discovered running processes, performance problems, deviations, and bottlenecks. Process owners should be able to take some actions based on this information with a certain level of confidence. To do so, they need to improve/change their processes interactively. Therefore, simulation and prediction techniques are taken into account to foresee the process after changes and improvement. Simulation techniques are capable of replaying processes with different scenarios.  

Process mining also enables designing data-driven simulation models of processes \cite{SummerSimKeynote2018}. However, in the current tools for simulation in process mining either interaction with the user and user knowledge is a prerequisite of designing a simulation model or the tools are highly dependent on the interaction of multiple simulation tools. In \cite{howcloseGawin}, an external tool, i.e., ADONIS for simulating the discovered model and parameters are used. 
The combination of BPMN and process mining is presented in \cite{DBLP:conf/bpm/CamargoDR19} in which BIMP is used as a simulation engine. However, the possibility of interaction for changing the process model is not available for the user. Also, the authors in \cite{Pufahl2017} propose a Java-based discrete event simulation of processes using BPMN models and user interaction where the user plays a fundamental role in designing the models.  
Generating a CPN model based on CPN tools \cite{DBLP:conf/apn/Westergaard13} is presented in \cite{DBLP:journals/is/RozinatMSA09}. The user needs to deal with the complexity of CPN tools and \emph{SML}. In \cite{netjes2010}, the focus is also to generate CPN models and measuring performance measurements using the \emph{Protos} models which can be used easier but more restricted than CPN tools.
 \cite{DBLP:journals/is/Rogge-SoltiW15} performs simulation on top of discovered Petri nets and measure the performing metrics and not re-generating the complete behavior of a process. 
The \emph{Monte Carlo} simulation technique, as well as generating sequences of activities based on process trees, are also proposed in Python \cite{DBLP:pm4py}. However, the simulation results are not in the form of an event log, 
and they lack the time perspective. Also, the tool in \cite{mahsaPetrinetToolSim} as a python library simulates the Petri nets of processes based on their event logs.
In \cite{MahsaBIS}, aggregated simulations which are useful for what-if analyses in high-level decision-making scenarios are introduced. The \emph{PMSD} tool represents the aggregated approach and generates a simulation model at a higher level of detail \cite{pourbafrani2020pmsd} based on the approach in  \cite{DBLP:conf/otm/PourbafraniZA19}. Different process variables are introduced in \cite{FeatureExtractionCaise21} that makes the different levels of process simulations possible, e.g., simulating the daily behavior of a process instead of simulating every event in the process.

\begin{figure}[bt]
    \centering
    \includegraphics[width=\textwidth,height=0.24\textheight ]{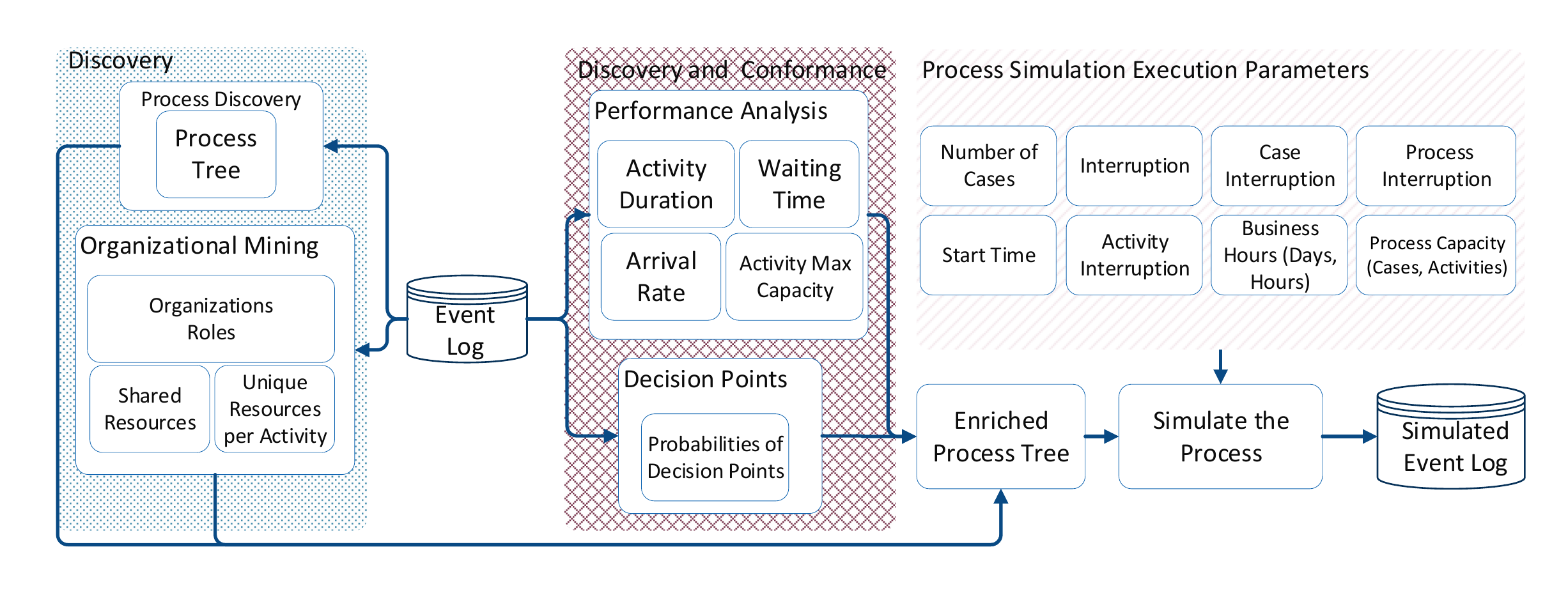}
        \caption{The general framework of our tool for generating a process tree using process mining techniques and enriching the process tree with the possible information from an event log. The resulting process model can be executed to re-run the process with the user-provided configuration. }
    \label{fig:Main_Framework_PMSIM}
\end{figure}

The interactive improvement/changes of processes is not a straightforward task when it comes to changing the process models and parameters by the user. Therefore, we use the discovered process tree and provide an interface for the user to discover and design a new process model including all the performance and environmental attributes, e.g., changing business hours, or resource capacity. All of these changes are supported by the information derived from event logs of processes using process mining techniques as shown in \autoref{fig:Main_Framework_PMSIM}. 
In this paper, we present our tool which is designed to support process improvement using simulation and process mining. Our tool is implemented as an integrated Python web-application using \emph{Django} framework, where all the modules are also accessible outside the user interface for the users to re-use or add their desired modification to the code. Moreover, to the best of our knowledge, it is the first tool that runs the possible changes in the process tree while considering performance aspects such as resource, activity, and process capacity, accurate time, as well as business hours. These capabilities are directly used to address interactive process discovery as well as process improvement.

\section{SIMPT}

As shown in \autoref{fig:Main_Framework_PMSIM}, our tool automatically extracts all the existing aspects of a process based on general event log attributes. These quantitative aspects are used for running the process tree w.r.t. different scenarios. The main aspect is the discovered process tree which is used for interaction with users since it is limited to $4$ operations. The user-interaction using process trees is easier compared to the complexity of Petri nets. 
The sequence, parallel, loop, and XOR operators are easy to be represented and understandable by the user for further changes. The process structure and the flow of activities are represented using these operators. 
We extend the implementations of the process tree method in \cite{DBLP:pm4py} to generate a comprehensive tree including the probability of choices (XOR) and loop restrictions, e.g., the maximum length of traces and execution of loops. 
For instance, measuring the performance KPIs of a process in case that activity \emph{a} and \emph{b} are parallel instead of being sequential is possible. 
Not only the possible traces are generated but also generating the complete event log gives all the performance aspects of the new process, e.g., the length of the queues, the service time of cases, and other possible aspects from an event log. 
\begin{table}[bt]
\caption{The general parameters in the tool are listed. Most can be derived using process mining. Also, the required execution parameters. All the parameters discovered using process mining from event logs by default and are filled in the tool with the real values automatically. The execution values guaranteed the default values in case that users do not change/provide the parameters. Note that the handover matrix is used for logging resources.  }
\label{table:param}
\resizebox{\textwidth}{!}{
\begin{tabular}{c|c|c|c|c|c|c|c|c|c|c|c|c|c|}
\cline{2-14}
 &
  \multicolumn{11}{c|}{Process Mining} &
  \multicolumn{2}{c|}{\begin{tabular}[c]{@{}c@{}}Simulation Execution\\ Parameters\end{tabular}} \\ \cline{2-14} 
 &
  \begin{tabular}[c]{@{}c@{}}Process\\  Model \\ (Tree)\end{tabular} &
  \begin{tabular}[c]{@{}c@{}}Arrival \\ Rate\end{tabular} &
  \begin{tabular}[c]{@{}c@{}}Activity\\  Duration,\\ Deviation\end{tabular} &
  \begin{tabular}[c]{@{}c@{}}Activities\\  Capacity\end{tabular} &
  \begin{tabular}[c]{@{}c@{}}Unique \\ Resources\\ (Shared\\  Resources)\end{tabular} &
  \begin{tabular}[c]{@{}c@{}}Social \\ Network \\ (Handover\\  Matrix)\end{tabular} &
  \begin{tabular}[c]{@{}c@{}}Waiting \\ Time\end{tabular} &
  \begin{tabular}[c]{@{}c@{}}Business \\ Hours\end{tabular} &
  \begin{tabular}[c]{@{}c@{}}Activity-flow \\ Probability\end{tabular} &
  \begin{tabular}[c]{@{}c@{}}Process\\  Capacity \\ (cases)\end{tabular} &
  \begin{tabular}[c]{@{}c@{}}Interruption \\ (Process, Cases,\\  Activities)\end{tabular} &
  \begin{tabular}[c]{@{}c@{}}Start Time \\ of Simulation\end{tabular} &
  \begin{tabular}[c]{@{}c@{}}Number\\  of Cases\end{tabular} \\ \hline
\multicolumn{1}{|c|}{\begin{tabular}[c]{@{}c@{}}Automatically \\ Discovered\end{tabular}} &
  + &
  + &
  + &
  + &
  + &
  + &
  + &
  + &
  + &
  + &
  + &
  - &
  - \\ \hline
\multicolumn{1}{|c|}{\begin{tabular}[c]{@{}c@{}}Changeable \\ by User\end{tabular}} &
  + &
  + &
  + &
  + &
  + &
  - &
  + &
  + &
  + &
  + &
  + &
  + &
  + \\ \hline
\end{tabular}
}
\end{table}
The provided insights also include the possibility of checking the newly generated behaviors by the new process structure (process tree) and configuration for conformance checking too.
\autoref{table:param} shows the general overview of the parameters and the possibility for the user to interact with the process and reproduce the new process behavior (event log) w.r.t. the changes in that parameters. Here, we explain some of the main modules. All the details along with the tool source code and a representing video are presented extensively. \footnote{https://github.com/mbafrani/SIMPT-SimulatingProcessTrees.git}

\begin{figure}[bt]
    \centering
    \includegraphics[width=\textwidth,height=0.25\textheight]{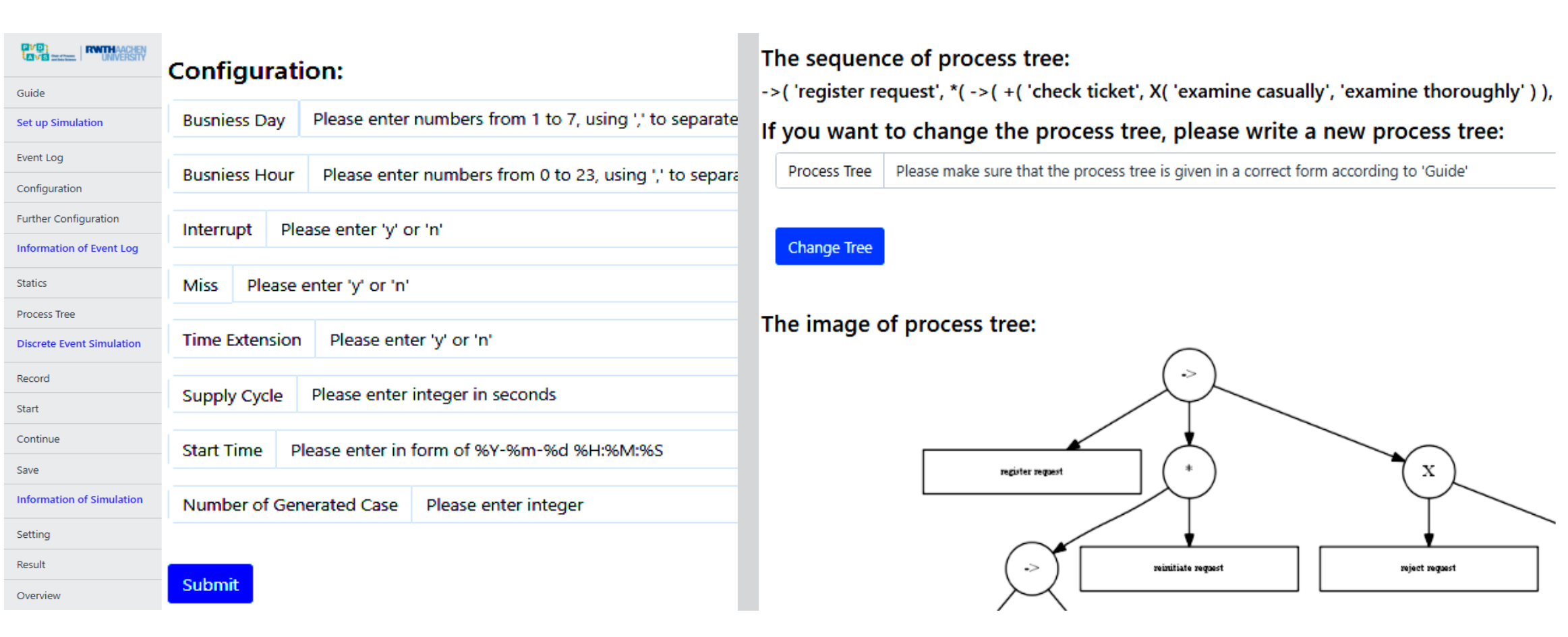}
    \caption{A part of the tool parameters configuration to apply the changes to the basic performance elements as well as the process tree structure tab for the possible changes. }
    \label{fig:appscreenshot}
\end{figure}

The tool has three main modules. The first module runs the discovery, extracts the current parameters in the event logs, and presents the model and the parameters to the user to decide on the possible changes, e.g., process tree, deviations, or waiting time. For the performance analysis, both event logs with start and complete timestamps and only one timestamp can be considered. The activities' durations are also taken from their real averages and deviations. 
The second module is configuring the new process parameters for simulating and running the simulation for the given parameters, e.g., the number of cases to be generated or the start time of the simulation. Furthermore, the interruption concept for process, activities, and traces is introduced. The user can define whether in specific circumstances, e.g., when a running case or activities is passing the business hours, the interruption can happen and it is logged for the user. The last module is running and simulating the defined and configure process tree in which the results are presented as an overview as well as the possibility for downloading as an event log. The handover matrix of the process is also used to log the resources based on reality in the generated event log.  The simulation is generating new events, e.g., the arrival of a new case or the start/end of an activity for a case, based on the clock of the system and configured properties, e.g., available resources. A part of the tool interface is shown in \autoref{fig:appscreenshot}, the \emph{Guide} section in the tool provides features, possibilities, required steps, and information, extensively.
The python library \emph{simpy} \footnote{simpy.readthedocs.io} is used for discrete event simulation and handles the required system clock to generate new events. 

\section{Tool Maturity}

The tool has been used in different projects to design/improve a process model interactively in different situations and generate different event logs. In the IoP project\footnote{www.iop.rwth-aachen.de}, the tool is used to simulate multiple production lines to estimate the effect of the capacity of activities on the production process, e.g., average production speed. Moreover, \cite{mahsaTimeseries} exploits the tool for car company production analyses, different arrival rates and activities’ duration for the same process has been selected and the tool event logs are generated.
Also, we use the tool as the base of the interactive design of the job-shop floor. The possible flow of jobs in the job-shop floor, i.e., the flow of activities, is presented as a process tree. These trees omit forbidden actions in the production line using the knowledge of the production manager and simulate the production line with the desired setting.

\begin{wrapfigure}{r}{0.5\textwidth}
     \vspace{-4 mm}
     \centering
    \includegraphics[width=0.5\textwidth,height=0.1\textheight ]{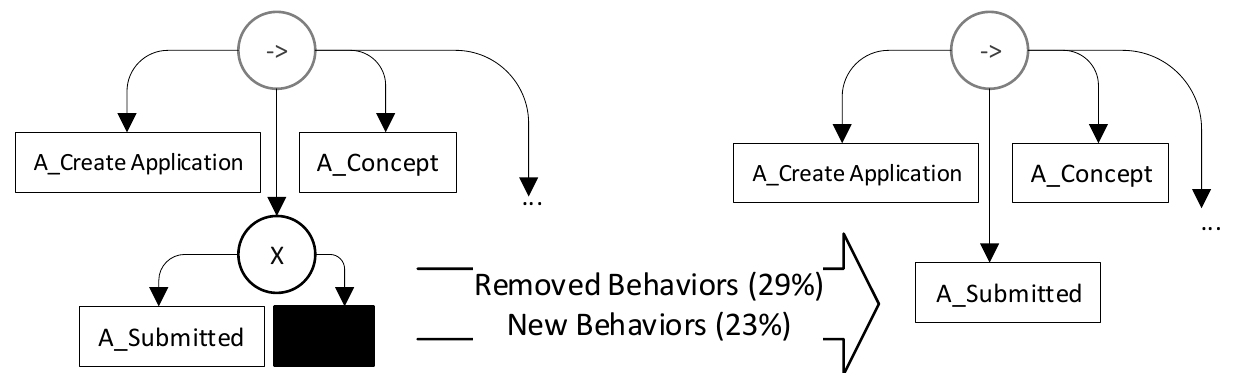}
    \caption{A sample scenario for the process model of the BPI Challenge 2017 event log (application requests) in the tool. Activity \emph{A-Create Application} can be skipped in the discovered process tree (left). By changing the choice to a sequence (right), i.e., this activity is required for all the cases, the removed and new inserted behaviors in the process can be measured. }
    \label{fig:sampleScnario}
    \vspace{-5 mm}
\end{wrapfigure}

\autoref{fig:sampleScnario} presents a sample scenario of changing the process structure and measuring the differences after the changes. Note that the inserted behaviors are generated based on the choices and loops in the rest of the process. 
As mentioned, having both simulated and original behavior of the process (with or without modifications) creates the possibility of the comparison between two processes which is available using the existing process mining tools and techniques. 
To demonstrate the tool functionality and validity of the re-generated event log, we used the BPI Challenge 2012 event log. We assessed the similarity of the original event log with the re-generated one using \emph{Earth-Mover Distance} (EMD) technique as presented in \cite{sanderEMD2019}
using the Python implementation in \cite{MajidQuantification}. EMD calculates the shortest distance between the two event logs w.r.t. the minimum movement over the minimum distance between traces.
The process is re-run without any changes in performance parameters to check the process behavior w.r.t. the flow of activities. The value of $0.34$ as the EMD measure indicates the similarity of the two event logs. Note that the choices in the process model are the reason to have more behavior than the real event log which is expected w.r.t. \emph{precision metrics} of discovery algorithm. Given the closeness of the simulation results and the original event log, the next changes can be applied to the process tree and other aspects of the process with enough confidence to reproduce the new event log including the effects of changes. 

\section{Conclusion}
Given the fact that process improvement is the next step in the process mining path, simulation techniques will become more important. 
The combination of process mining and simulation techniques makes it possible to improve and re-design the processes w.r.t. the discovered possible change points in processes. Our tool is designed and implemented with the purpose of making process improvement using process mining and simulating the processes with the user's changes possible. The process tree notation along with all the performance and execution aspects of the process make the generated new behavior of the process w.r.t. user possible improvement reliable. Based on the provided reliable platform, the possibility of recommending the best process model interactively with the user considering both performance and activity flow is the next step.

\bibliographystyle{splncs04}
\bibliography{Reference}
\end{document}